\documentclass[aip,graphicx]{revtex4-1}

\usepackage{amsmath,color,graphicx}
\newcommand{\cc}[1]{\textcolor{black}{#1}}

\draft 

\begin{document}

\title[Liquids in nanopores]{Perspective Article: Liquids confined in hydrophobic nanopores}
\title[Boiling in nanopores]{What keeps nanopores boiling}

\author{Alberto Giacomello}
 \email{alberto.giacomello@uniroma1.it.}

\affiliation{Dipartimento di Ingegneria Meccanica e Aerospaziale,
Sapienza Universit\`a di Roma, 00184 Rome, Italy}%

\date{\today}

\begin{abstract}
The liquid to vapour transition can occur at unexpected conditions in nanopores, opening the door to fundamental questions and new technologies. The physics of boiling in confinement is progressively introduced, starting from classical nucleation theory, passing through nanoscale effects, and terminating to the material and external parameters which affect the boiling conditions. The relevance of boiling in specific nanoconfined systems is discussed, focusing on heterogeneous lyophobic systems, chromatographic columns, and ion channels. The current level of control of boiling in nanopores enabled by microporous materials, as metal organic frameworks, and biological nanopores paves the way to thrilling theoretical challenges and to new technological opportunities in the fields of energy, neuromorphic computing, and sensing. 
\end{abstract}


\maketitle 

\section{Introduction}

Liquids in narrow confines are nothing like bulk ones, leading to exotic phase behaviours \cite{huber2015} and transport properties \cite{kavokine2021} that unlock novel applications \cite{robin2021,you2022}. For example, water in zeolites boils at extremely large pressures\cite{eroshenko2001} (100 MPa at 20$\,^\circ$C), giant slip is observed in carbon nanotubes \cite{secchi2016}, and selectivity towards specific ions \cite{doyle1998} is achieved by biological ion channels with conductivity near the diffusion limit \cite{horn2014}. 
The focus of this Perspective is boiling --a term which is used broadly, meaning the liquid to vapour transition, also referred to as evaporation, cavitation, or drying in the literature-- of liquids in \cc{repulsive} nanopores: the conditions at which it happens, the properties that follow, and the fundamental and technological fields in which this phenomenology matters. Nanoscale confinement \cc{together with repulsive solid-liquid interactions, which make the pores hydrophobic in the case in which water is the liquid of interest, or more generally solvophobic/lyophobic,} changes the face of boiling, defying common intuition: it can occur at extremely high pressures \cite{eroshenko2001} or at low temperatures \cite{chorazewski2021}, as opposed to the usual $100\,^\circ$C at $1$~atm for water, it can be controlled by acting on the geometry and chemistry of the nanopores \cite{giacomello2020}, on the liquid \cite{ortiz2014,grosu2018}, or on external fields \cite{trick2017}; it can couple to the flexibility of the porous matrix \cite{lowe2019}, giving rise to negative compressibility \cite{tortora2021} or be instrumental for switching ionic currents in neurons  \cite{roth2008,aryal2015}. 
The topic of liquid behaviour at the nanoscale is much broader than what can be touched upon here; the interested readers are referred to recent reviews for additional insights \cite{huber2015,kavokine2021,coudert2021,giacomello2021,you2022,le_donne2022,robin2023b}.

As any phase transitions, boiling involves the formation of a new phase and the competition of bulk gains and surface costs typical of nucleation phenomena \cite{volmer1939,kelton2010}.
In heterogeneous nucleation, the presence of a solid wall can actually decrease the nucleation cost, even providing an energy gain in the case of lyophobic surfaces \cite{skripov1972}.
In nanopores, heterogeneous nucleation is brought to its extreme, with confinement conspiring with lyophobicity to thermodynamically favour the vapour phase and substantially decrease the nucleation barriers \cite{giacomello2020}. Nanosystems across different realms, ranging from engineering to biology and soft matter, present suitable conditions to significantly alter boiling properties: hydrophobic nano- and microporous materials \cite{eroshenko2001,lefevre2004,grosu2016}, biological ion channels \cite{aryal2015,rao2018} and nanopores \cite{lucas2021,paulo2023c}, solid-state nanopores \cite{smeets2006b,smeets2008}, and several others which are likely to emerge as our knowledge and manipulation capabilities at the nanoscale increase.

Nanoconfinement-enhanced boiling has been reported in an increasing number of systems; three will be considered in some detail: heterogeneous lyophobic systems (HLS), the stationary phases of high performance liquid chromatography (HPLC), and biological ion channels.
HLS are constituted by water (or another non-wetting liquid) and hydrophobic (or lyophobic) nanoporous materials; leveraging their enormous surface area per unit mass and the transitions between confined vapour and liquid, HLS allow to dissipate or store energy \cite{eroshenko2001} in an extremely compact way. Although the focus is on boiling, which is a less investigated and more elusive confined phase transition,  the vapour to liquid transition (``intrusion'') will also be discussed, due to its crucial role in the \cc{typical operation of HLS, which  consists of the successive hydrostatic compression and decompression of the system,} see Fig.~\ref{fig:int-ext}a. 
Some of the hydrophobic nanoporous materials used for HLS are also employed as the stationary phase in reversed-phase HPLC, an important separation technique, which indeed may be subject to ``retention lossess'' due to the boiling (or dewetting) of highly aqueous solvents confined in the nanopores  \cite{walter1997,walter2005,gritti2019,gritti2020}.
Finally, a phenomenology analogous to boiling occurs in some biological ion channels (and other nanopores) -- pore-forming proteins which allow the controlled transport of water and solutes across the hydrophobic environment of the cellular membrane. It appears that the presence of hydrophobic motifs can control the formation of bubbles in biological nanopores \cite{beckstein2003,anishkin2004,roth2008,aryal2015,rao2018}, which in turn block the transport of ions \cite{zhu2012a}, see Fig.~\ref{fig:ionchannel}a-b. This mechanism is known as hydrophobic or bubble gating and allows ion channels to switch ionic currents even when there is no steric block of the pore \cite{rao2019}, at a small energetic cost \cite{roth2008}.

The three examples above illustrate well the program of this Perspective: understanding and, eventually, controlling the conditions at which boiling occurs in nanopores to unlock new technological opportunities. 
On the fundamental side, the investigation of boiling in nanopores opens new routes to study nanoscale quantities, such as line tension \cite{guillemot2012a,tinti2017}, and biologically relevant phenomena such as hydrophobic gating 
\cite{roth2008,aryal2015,giacomello2020}. On the technological side, finding the design and control parameters that govern boiling can enable and reinforce energy applications of HLS \cite{grosu2018}, improve chromatographic columns \cite{gritti2020}, and lead to novel bioinspired applications of nanopores \cite{robin2023b,reitemeier2023}.
Substantial challenges are in the way of this manifesto, which makes it attractive and yet unaccomplished. Foremost is the multiscale nature of boiling in nanopores: this unexpected behaviour has nanoscale origin and macroscopic consequences. In experiments, direct measurement of microscopic mechanisms is often impossible, calling for new approaches \cite{guillemot2012b,tortora2021,gritti2019} and for the support of theoretical or computational models \cite{coudert2021}; even structural information at the (sub)nanoscale may be arduous to obtain \cite{doyle1998}.  Nanoscale phenomena, however, often defy our understanding based on macroscopic theories, e.g., relevant thermodynamic quantities or classical nucleation theory fails at describing them \cite{helmy2005}. While simulations promise to provide interpretation to experiments and bridge the diverse scales, the computational burden of dealing with  multiple lengths and  times poses significant challenges
\cite{meloni2016,giacomello2021}. 

This Perspective attempts to gradually introduce the physical concepts needed to understand boiling in confinement, progressively increasing the complexity of the systems and phenomena considered (Sec.~\ref{sec:phase}). Subsequently, selected cases are discussed  in which boiling is important (Sec.~\ref{sec:impact}). Section~\ref{sec:perspectives} discusses open issues and future perspectives in the field, while Sec.~\ref{sec:conclusions} is left for conclusions.

\section{Reaching boiling point in confinement}
\label{sec:phase}
\subsection{Fundamentals}
\label{sec:fundamentals}
Nucleation phenomena, as boiling, are classically understood as a competition of bulk terms, which tend to favour the nucleating phase over the metastable one, and surface terms, which constitute the energetic cost to form the new phase; the different scaling of these terms with \cc{the size of the nucleus} leads to the appearance of an energy barrier which can be crossed by random thermal fluctuation \cite{volmer1939,skripov1972}.
This simple picture also holds for boiling in nanopores, but the role of surface terms is more subtle, as there is a significant contribution of the confining surfaces, as seen by writing the free energy of a two-phase system in contact with a solid wall \cite{giacomello2020}:
\begin{subequations} 
\label{eq:FE}
\begin{equation}
\Omega = -P_l V_l -P_vV_v + \gamma_{lv} A_{lv} + \gamma_{sv} A_{sv}+\gamma_{sl} A_{sl}, \text{ or}
\end{equation}
\begin{equation}
\label{eq:FE2}
\Delta \Omega \equiv \Omega- \Omega_\mathrm{ref}=\Delta P V_v + \gamma_{lv} \left(A_{lv} + \cos\theta_Y A_{sv}\right) ,
\end{equation}
\end{subequations}
where the subscripts $l$, $v$, and $s$ denote the liquid, vapour, and solid phases, respectively, and the related interfaces; $P$ are the bulk pressures, $\gamma$ the surface tensions, $V$ the volumes, and $A$ the surface areas. To obtain eq.~\eqref{eq:FE2}, the total  volume of the system and the total  surface area of the solid are assumed to be constant, i.e., $V_l=V_{tot}-V_{v}$ and $A_{sl}=A_{tot}-A_{sv}$, respectively; Young's equation $\gamma_{lv}\cos\theta_Y=\gamma_{sv}-\gamma_{sl}$, with $\theta_Y$ Young's contact angle, and $\Delta P\equiv P_l -P_v$  were also used. 
Equation~\eqref{eq:FE2} groups  together the constant terms in $\Omega_\mathrm{ref}=-P_lV_{tot}+\gamma_{sl}A_{tot}$, which is the reference free energy of the confined liquid.
In bulk nucleation $A_{sv}=A_{sl}=0$. In confinement, the additional term related to the wall can have either positive or negative sign for  lyophilic ($\theta_Y<90^\circ$) or lyophobic walls ($\theta_Y>90^\circ$), respectively. This means that boiling can be  favoured by confinement opening new possibilities to control boiling.

\begin{figure}
    \centering
    \includegraphics[width=0.7\textwidth]{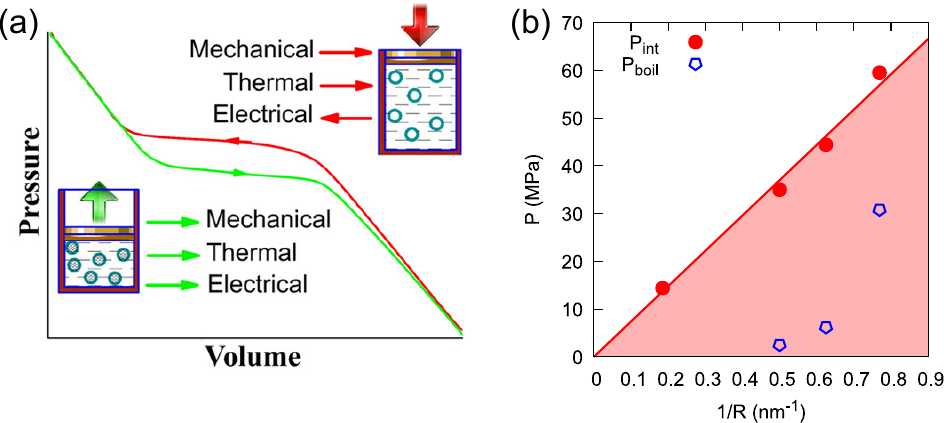}
    \caption{a) Typical pressure vs volume diagram during a compression (red) / decompression  (green) experiment in HLS composed of a liquid and a lyophobic nanoporous material sealed in a container whose volume can be controlled. The plateau at higher pressure corresponds to the intrusion process while the one at lower pressures to boiling. The possible forms and typical signs of the energy exchange with the environment are indicated by arrows. Reprinted with permission from Grosu et al. ACS Appl. Mater. Interfaces 9, 7044 (2017). Copyright 2017, American Chemical Society.
    b) Intrusion and boiling pressures in silanised MCM-41 of different radii. The intrusion pressures (red circles) are well fit by a straight line corresponding to Kelvin-Laplace eq.~\eqref{eq:lcap} with $\theta_Y=120.9\,^\circ$. The red shaded area indicates the region in which the capillary vapour is expected to be stable while the liquid metastable. The boiling pressures (blue pentagons) all fall in the deep metastable liquid region. Data from Lefevre et al.\cite{lefevre2004}. }
    \label{fig:int-ext}
\end{figure}
For the special case of an infinite cylindrical capillary of radius $R$ occupied by vapour only ($A_{lv}=0$), eq.~\eqref{eq:FE2} allows to find the conditions at which the confined vapour and liquid
coexist ($\Omega=\Omega_\mathrm{ref}$), which is known as Kelvin-Laplace equation:
\begin{equation} \label{eq:lcap}
  \Delta P =- 2 \frac{\gamma_{lv} \cos \theta_Y}{R}.
\end{equation}
Although strictly valid only for the coexistence of capillary liquid and vapour in infinite cylindrical pores and in equilibrium (infinite times), this equation introduces the two main actors of the thermodynamics of boiling in nanopores: surface lyophobicity ($\theta_Y$) and characteristic size of the confinement ($R$). At constant temperature, in the presence of lyophobic walls ($\cos\theta_Y<0$), boiling can occur at pressures much higher than the coexistence one ($\Delta P\gg 0$). The pressure deviation $\Delta P$ becomes significant in nanopores, i.e., when $R$ is sufficiently small: for water, $R=2$~nm, and $\theta_Y=110^\circ$, boiling occurs at pressures as large as $\Delta P =25$~MPa at ambient temperature. 

\cc{Figure~\ref{fig:int-ext}b shows that, for long, cylindrical nanopores (hydrophobised MCM-41), Kelvin-Laplace equation~\eqref{eq:lcap} describes well the intrusion pressure and its $1/R$ dependence, while it fails to render the boiling pressure. In principle, both phenomena are nucleation events which occur between the coexistence and the spinodal pressures. For intrusion in long cylindrical nanopores, the two pressure are indistinguishable \cite{rasera2023}, resulting in the match of Fig.~\ref{fig:int-ext}b. On the other hand, the boiling spinodal pressure occurs far from coexistence down to negative pressures \cite{tinti2017}, which allows boiling to occur over a much broader range of pressures (blue pentagons in Fig.~\ref{fig:int-ext}b). Accordingly, it is easier to distinguish in boiling the characteristics of nucleation described in Sec.~\ref{sec:extrinsic}, including a marked dependence of the boiling pressure on temperature and on observation time. 
}

Equation~\eqref{eq:FE2} also implies that the kinetics of nucleation are accelerated by the presence of lyophobic walls. The classical argument \cite{skripov1972} is that the critical bubble (maximum of the free energy determining the nucleation barrier) will be a spherical cap meeting the wall with the prescribed contact angle, which decreases the barrier down to zero for $\theta_Y=180^\circ$; curvature has a similar effect, with concave walls decreasing the barrier \cite{giacomello2020}. In nanopores, the nucleation path may not correspond anymore to a sequence of spherical caps valid in the bulk or at nearly flat surfaces \cite{meloni2016}, because the size of the bubble becomes comparable to the confinement; the critical bubble for cylindrical pores resembles a saddle  \cite{lefevre2004,tinti2017,tinti2018}, which limits its variability in terms of volume \cite{guillemot2012a} and further accelerates the boiling kinetics \cite{tinti2017}.

Boiling is indeed observed in nanopores at large pressures: Sec.~\ref{sec:impact} discusses selected examples in nanoporous materials, HPLC columns, and biological nanopores. The interested reader may also refer to the literature about cavitation in confinement \cite{berard1993,bolhuis2000,duan2012,ashbaugh2013,giacomello2013,amabili2016b}, capillary evaporation \cite{lum1999,luzar2000,truskett2001,leung2003,roth2006,roth2011}, extrusion \cite{lefevre2004,grosu2014,khay2014} or dewetting \cite{huang2003,powell2011,remsing2015,gritti2019} from nanopores, which all boils down to the same phenomena.

\subsection{Nanoscale effects}
\label{sec:nanoscale}
In nanopores the macroscopic free energy in eq.~\eqref{eq:FE} may fail to be predictive because nanoscale effects come into play. An important one is the free energy cost to bend solid-fluid interfaces, which can be accounted for by adding bending rigidities terms to eq.~\eqref{eq:FE2}, as done, e.g., in morphometric thermodynamics \cite{roth2006}. Similarly, curved liquid-vapour interfaces may introduce measurable corrections to nucleation kinetics \cite{bruot2016}. 
Furthermore, the presence of three-phase contact lines and the related thermodynamic \cc{force}, line tension  \cite{schimmele2007}, can play an important role in nanoconfined boiling. These line effects have been reported in simulations of water confined by parallel hydrophobic plates at nanoscale separations \cite{sharma2012a,remsing2015,tinti2023}, but is further enhanced by the curved surface of nanopores \cite{tinti2017}. Indeed, experiments on hydrophobised silica gels demonstrated that the large boiling pressures could be explained only by line tension with negative sign, i.e., facilitating boiling \cite{guillemot2012a}. The experimental value of line tension, which is system and definition  dependent\cite{schimmele2007}, was ca. $-24$~pN for MCM-41. Later, molecular dynamics simulations  \cite{tinti2017,paulo2023a} on similar hydrophobic nanopores obtained an estimate of $-10$~pN, demonstrating that line tension is capable of reducing the boiling free-energy barrier by a factor five. 

Below the capillary critical point \cite{evans1990}, the boiling transition is first order, which implies hysteresis phenomena over the metastable range. Indeed a clear hysteresis loop is observed in the typical $P$-$V$ isotherms which are used in the standard tests and applications of HLS, see, e.g., Fig.~\ref{fig:int-ext}a; the process opposite to boiling 
is known as ``intrusion''  in the HLS literature and occurs at higher pressures, $P_\mathrm{int}>P_\mathrm{boil}$, see Fig.~\ref{fig:int-ext}b. Controlling such hysteresis is one of the key challenges for engineering HLS with controlled energy dissipation or storage properties. When the nanopores are particularly narrow or lyophobic, hysteresis can be drastically reduced \cite{tinti2017} and one could anticipate intermittent filling: this was reported in simulations of water in carbon nanotubes  \cite{hummer2001,waghe2002}, model nanopores \cite{allen2002,allen2003,beckstein2003}, and biological ion channels \cite{anishkin2004,beckstein2006}.

\subsection{Influence of nanoporous material characteristics on boiling}
\label{sec:intrinsic}
Subtle pore characteristics may have an influence on boiling, beyond the simple picture of Sec.~\ref{sec:fundamentals}, which just invokes size and hydrophobicity, and beyond the nanoscale effects of Sec.~\ref{sec:nanoscale}. Understanding these effects is key to design new systems with better control over boiling.
A paradigmatic example is the case of two hydrophobised silica gels with comparable pore sizes (between $6$ and $9$~nm), which exhibited qualitatively different boiling behaviour \cite{amabili2019}: the material with independent pores showed no boiling down to ambient pressures, while the one with interconnected pores displayed boiling at $2$~MPa. This unexpected difference could be linked by molecular dynamics and theory to the presence of nanometre-sized interconnections between pores which were always empty at ambient pressure, making the surface of the  main pores effectively superhydrophobic; on the other hand, independent pores were too large to allow boiling at ambient conditions \cite{amabili2019}.

Actual lyophobic surfaces always have some degree of heterogeneity \cite{fraux2017}. Indeed, in nanopores even molecular heterogeneities can play a role in the kinetics of boiling \cite{luzar2000}, which is rooted in the capability of nanodefects to pin the liquid-vapour interface \cite{giacomello2016a}. Pores with controlled hydrophilic/hydrophobic nanometre-sized patches (periodic mesoporous organosilicas) showed the logarithmic signature of thermally activated jumps of the liquid meniscus over nanoscale anchoring defects during the intrusion process \cite{picard2021}, although the boiling kinetics was governed by vapour nucleation as in simple pores \cite{lefevre2004}. 
Recent molecular dynamics simulations of nanopores   functionalised with hydrophobic chains  showed that the random heterogeneities which emerge at different grafting densities and chain lengths play a crucial role both in intrusion and in boiling\cite{rasera2023}. Local defects in the grafting may pin the interface and increase the intrusion pressures, while serving as nucleation seeds that facilitate boiling. In short, very local surface characteristics may govern intrusion and boiling in nanopores beyond the simple intuition based on eq.~\eqref{eq:lcap} that the pressure of the liquid to vapour transition should only depend on hydrophobicity and size.

\begin{figure}
    \centering
    \includegraphics[width=0.9\textwidth]{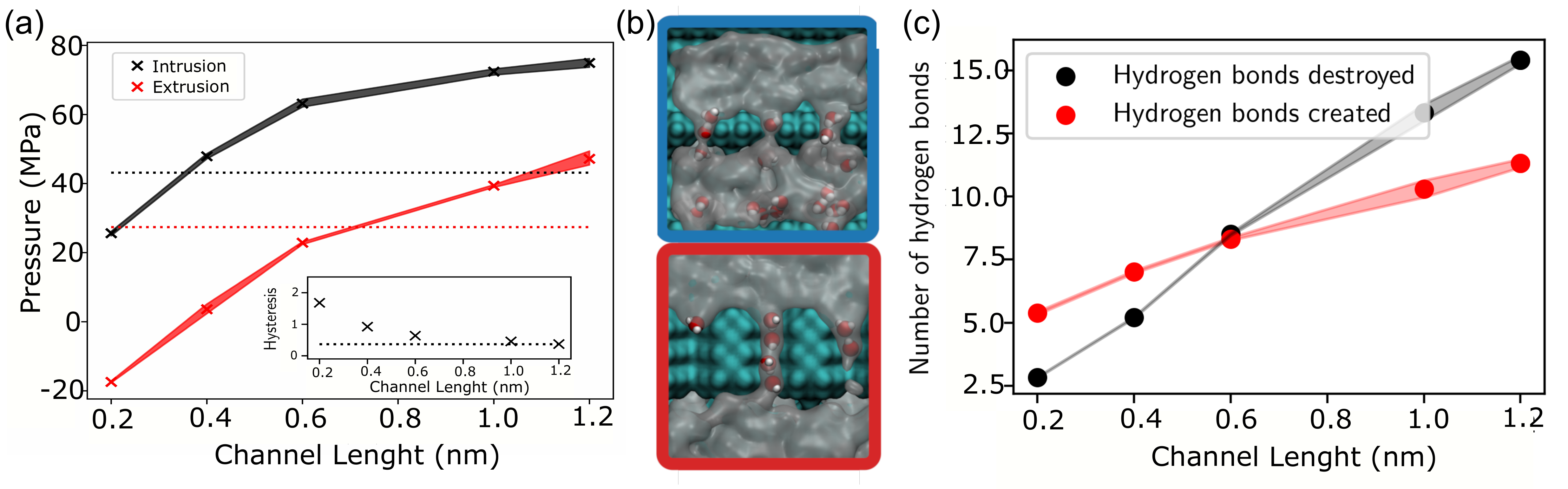}
\caption{a) Intrusion (black) and boiling (red) pressures in cylindrical nanopores with diameter $1.54$~nm as a function of the length of the secondary channels  with diameter $0.77$~nm connecting them; dotted lines report the same quantities for a cylindrical pore without secondary channels. 
b) Formation of a single-file arrangement of water molecules inside secondary channels of length $0.2$~nm (blue) and $1.0$~nm (red).
c) Number of hydrogen bond created (red) and destroyed (black) by moving a water molecule from the cylindrical pore to the secondary channels of different lengths. 
 Adapted from Paulo et al., Comm. Phys. 6, 21 (2023). Copyright 2023 Authors.\cite{paulo2023b}.}
    \label{fig:nanoscale}
\end{figure}

With the emergence of hydrophobic microporous materials with a well defined crystalline structure \cite{soulard2004,fraux2017}, the important aspects discussed above of pore connectivity and heterogeneity can be controlled with molecular precision. The size close to or smaller than the nanometre, however, poses new challenges to the understanding of capillary phenomena, beyond the nanoscale effects discussed in Sec.~\ref{sec:nanoscale}; such angstroscale effects are touched upon in Sec.~\ref{sec:perspectives}. As an example, it was shown that, unlike in silica gels \cite{amabili2019}, the presence of subnanometric secondary channels between main pores in model zeolites can facilitate intrusion of water and disfavour boiling \cite{bushuev2022}; the role of hydrogen bonding bridging the main pores across the secondary channels was anticipated. Indeed, systematically changing the length of the secondary channels showed that the effective hydrophobicity of the main pores can change all the way from more hydrophilic to more hydrophobic than an independent pore \cite{paulo2023b}, i.e., disfavour or favour boiling, depending on whether the channels are long or short, respectively (Fig.~\ref{fig:nanoscale}a). Differently from the macroscopic expectations (eq.~\eqref{eq:lcap} would predict $\Delta P>100$~MPa), water was able to enter subnanoscale hydrophobic channels at ambient pressure. The origin of such behaviour could be understood considering the single file arrangement of water molecules within subnanochannels (Fig.~\ref{fig:nanoscale}b), which are able to avoid \cc{the formation of} two energetically costly dangling hydrogen bonds at the ends of the subnanometric cavity; the overall balance of created/destroyed hydrogen bonds is favourable when the subnanometric channels are shorter than a threshold \cite{paulo2023b} (Fig.~\ref{fig:nanoscale}c).  
Interestingly, the different topologies of micropores  (1D, 2D/3D, or cage-like) yield different trends of the intrusion pressure as a function of the accessible area to volume ratio \cite{bushuev2022b} providing new routes to control the phase behaviour of water \cite{coudert2021}; this may also have to do with the presence of subnanoscale connecting channels, but the mechanisms remains to be fully explained. 

The flexibility of confining surfaces is known to impact the boiling conditions, typically favouring it \cite{altabet2017} because it can decrease the volume of the critical bubble \cite{chorazewski2021}. This scenario is particularly relevant for flexible microporous materials, such as ZIF-8 or Cu$_2$(tebpz) \cite{grosu2016}, which have been shown to have a pronounced dependence \cc{on the compression/decompression rates} \cite{lowe2019,sun2021}. Flexibility of microporous materials can be also exploited to obtain (volumetric) negative compressibility
\cite{zajdel2021,tortora2021,caprini2023} (Fig.~\ref{fig:NC}a-f), which is present to a lesser degree even in mesoporous materials \cite{michel2022}.

The structure of microporous materials is defined at a molecular level, but these materials are not exempt from defects \cite{trzpit2008b,moslein2022} or finite size effects.
For the latter, it is known that the size \cc{of crystallites, i.e., of regions of regular crystalline structure, has} a significant influence on all variables of interest for HLS, including the boiling pressure  \cite{khay2015,sun2021,zajdel2022}. Recently, it was shown that the ZIF-8 half-cages at the crystallite surface are always occupied by water, which introduces effects which depend on the surface/volume fraction\cite{johnson2023}: this is obvious for the intruded volume, which is less in smaller crystallites, and less trivial for intrusion and boiling, which are made respectively easier and harder in nanoZIF-8. 
For intrusion, the surface half-cages favour a wetting mechanism which proceeds by the advancement of a coherent liquid front
\cite{amayuelas2023}, while, for boiling, the same phenomena discourages vapour nucleation at the surface.

\subsection{Influence of external parameters and fluid characteristics on boiling}
\label{sec:extrinsic}

While in Sec.~\ref{sec:intrinsic} the intrinsic nanopore characteristics which influence boiling were discussed, this section  focuses on the effects of external parameters and fluid characteristics. Such parameters provide additional knobs to control systems of technological interest or to understand fundamental biological phenomena.

In Sec.~\ref{sec:fundamentals} boiling in nanopores was introduced within the framework of nucleation, which explains why the pressure $P_\mathrm{boil}$ depends on temperature. If one assumes an Arrhenius law, the nucleation time 
\begin{equation}
    \label{eq:arrhenius}
    t = t_0 \exp \left( \frac{\Delta \Omega^\dag (\Delta P, \gamma, \theta_Y, \dots)}{k_BT}\right) \text{ ,}
\end{equation}
depends linearly on a prefactor $t_0$ and exponentially on the free-energy barrier $\Delta \Omega^\dag$, which could be computed based on eq.~\eqref{eq:FE2}. \cc{In eq.~\eqref{eq:arrhenius},  $k_B$ is the Boltzmann constant and $T$ the temperature.} One could invert eq.~\eqref{eq:arrhenius} to yield the boiling pressure $P_\mathrm{boil}$ ($P_v\approx 0$ for simplicity) as a function of the imposed nucleation time imposed by \cc{the experimental compression time}\cite{guillemot2012a}:
\begin{equation}
\label{eq:pext}
    P_\mathrm{boil} = \frac{k_BT}{V_c}\ln{\frac{t}{t_0}} + P_{0,\mathrm{boil}}(T)
\end{equation}
where $V_c$ is the volume of the critical bubble, and $P_{0,\mathrm{boil}}$ a reference boiling pressure at some chosen conditions. Equation~\eqref{eq:pext} suggests that $P_\mathrm{boil}$ should increase with temperature, which is indeed observed in experiments \cite{guillemot2012a,picard2021}
and simulations \cite{paulo2023a}. However, the dependence of $V_c$, $t_0$, and $P_{0,\mathrm{boil}}$ on temperature is much less clear and could introduce non-trivial effects, still to be fully explored. For example, it was shown\cite{grosu2018} that the dynamic viscosity $\eta$ enters $t_0$ introducing a logarithmic dependence $P_\mathrm{boil}V_c\propto - k_B T\ln \eta$; since $\eta(T)$ this also has an effect on the temperature dependence $P_\mathrm{boil}(T)$.

Equation~\eqref{eq:pext} also predicts a logarithmic dependence of $P_\mathrm{boil}$ on the experimental time of the decompression experiment, which is indeed observed in experimental data  spanning several decades \cite{guillemot2012b,guillemot2012a,picard2021}.  
This signature is seen to a lesser degree also in the intrusion process \cite{picard2021}; together these results suggest that the energy dissipation characteristic of HLS is only weakly time-dependent \cite{tinti2017}, which is a desirable characteristic in vibration damping applications. Moreover, the enhancement of hysteresis at low times, magnified by the flexibility of some microporous materials \cite{lowe2019}, may be exploited to effectively dissipate shocks and impacts \cite{sun2021}.

It is well known that the presence of gasses dissolved in the liquid can facilitate cavitation, by providing nuclei to initiate the process \cite{bussonniere2020}.
In nanopores, dissolved gases have indeed been suggested to facilitate boiling
\cite{li2019,camisasca2020,binyaminov2023}. The mechanism seems to be twofold: poorly soluble species accumulate at walls \cite{tortora2020}, especially concave ones, and reduce the nucleation barrier by enhancing density fluctuations in the pore \cite{camisasca2020}.
Such phenomenon has been hypothesised to play a role in general anaesthesia by volatile substances, by enhancing the hydrophobic gating process in some ion channels
\cite{roth2008}; in physical terms, this simply means that boiling is facilitated by the presence of poorly soluble gases in the bloodstream.

Electrolytes are used in HLS as means to increase the stored or dissipated energy because they increase both the intrusion and the boiling pressures 
\cite{khay2014,ortiz2014,ryzhikov2018,fraux2019,confalonieri2020}. For boiling, it is well known that solutions have a higher boiling point due to the reduction in the vapour pressure brought about by the solutes (boiling-point elevation). In nanopores, other properties affected by the presence of salts in the solvent can play a role, e.g., changes in viscosity and surface tension. For ZIF-8, it has been shown that some salts are rejected by the microporous material which acts as a sieve generating an increase in the intrusion and boiling pressures equal to the van 't Hoff osmotic pressure
\cite{michelin_jamois2015}. 
Electrolytes are also crucial for the very function of ion channels; even though the physiological concentrations are much smaller than in the HLS experiments above, the local one within the pore can be much higher. One may thus expect local salt concentration to play a role in those cases in which hydrophobic gating is relevant, but this remains to be investigated.

The presence of an electric field can further affect boiling in nanopores. Hansen and coworkers observed a that ion conduction is possible, i.e., the nanopore is wet, when the concentration gradient across a hydrophobic nanopore is high, while boiling occurs when the electric field is reduced, thus blocking further ion transport \cite{dzubiella2004}; in subsequent work, the group provided details on the electrostriction that water undergoes in the nanopore, which causes an increase in the density and a distortion of the hydrogen bond network \cite{dzubiella2005}; this was confirmed by later work \cite{bratko2007}.
This phenomenology, which is related to electrowetting at the nanoscale \cite{daub2007}, affords control on boiling in nanopores \cite{vanzo2015}, opening the way to realise voltage-gated hydrophobic nanopores
\cite{trick2017,smirnov2011,powell2011} (Fig.~\ref{fig:ionchannel}c), which are considerably simpler both in chemistry and gating mechanism than the corresponding ion channels.  Hydrophobically gated  nanopores display the electric characteristic of a memristor, i.e., a resistor with memory \cite{chua1976}, which is why they have recently been proposed as the basic element of nanofluidic neuromorphic computing architectures \cite{paulo2023c}.

\section{Where boiling in nanopores matters}
\label{sec:impact}
\subsection{HLS as energy materials}
\label{sec:energy}
Already in the 1980s Eroshenko \cite{eroshenko1982,eroshenko1988} foresaw the potential of hydrophobic nanoporous materials in the field of energy storage and dissipation. These HLS applications exploit 1) the very large specific surface areas (up to thousands of square meters per gram for microporous materials \cite{chae2004}) and 2) the tunable and often reversible confined phase transitions (intrusion and boiling). \cc{HLS are typically operated by hydrostatically compressing the system until intrusion occurs and subsequently decompressing it (Fig.~\ref{fig:int-ext}a).
}
Depending on the conditions at which intrusion and boiling occur HLS find application as \cite{eroshenko2000,eroshenko2001}: energy dampers (when the cycle is reversible and has large hysteresis), single-use bumpers (when intrusion is irreversible with large hysteresis), or energy storage devices (when intrusion and boiling occur at comparable conditions, leading to small hysteresis). It is thus apparent that controlling the intrinsic (Sec.~\ref{sec:intrinsic}) and extrinsic (Sec.~\ref{sec:extrinsic}) parameters which determine the boiling conditions is crucial for energy applications.

In the previous sections it was briefly mentioned that different classes of materials have been used as HLS. The oldest and still broadly adopted one is that of mesoporous silica gels, i.e., with pore size larger than ca. $2$~nm, functionalised with hydrophobic chains, e.g., by silanisation. Silica gels are cheap, can be mass-produced, and have remarkable stability; these characteristics allowed them to undergo advanced technological development, e.g., for automotive applications \cite{eroshenko2007,iwatsubo2007} passing  endurance tests \cite{suciu2008}. A variety of pore shapes are available, ranging from the almost ideally cylindrical MCM-41, which has been used extensively in theoretical studies \cite{lefevre2004,guillemot2012a}, to random interconnected ones like WC8 \cite{grosu2014}. As mentioned in Sec.~\ref{sec:intrinsic}, the connectivity of mesopores can be exploited to control boiling \cite{amabili2019}. While there is some control over the process and type of surface functionalisation \cite{fadeev1997,picard2021}, local defects, which are often random,  can have a significant impact on the performance of mesoporous HLS \cite{rasera2023}. Overall, while mesoporous silica are good materials to realise affordable and stable HLS, they have limited specific surface areas and \cc{restricted} control of the local wetting properties.

Ordered microporous materials \cite{davis2002}, such as zeolites, metal organic frameworks (MOFs), and covalent organic frameworks (COFs) are emerging in HLS applications \cite{fraux2017} because they promise to considerably increase the specific surface areas and enable molecular level control of the geometrical and chemical characteristics, unlike mesoporous silica.
Unfortunately, very few hydrophobic microporous material show sufficient stability for repeated use: the Silicalite-1 zeolite, which was among the first microporous materials to be used in HLS applications \cite{eroshenko2001,trzpit2008b,ryzhikov2018,khay2014}, is not stable under repeated compression/decompression cycles \cite{ievtushenko2013}. Concerning MOFs, ZIF-8 \cite{ortiz2013,grosu2015a,grosu2015b,khay2015,grosu2017,sun2018,sun2021,zajdel2022,amayuelas2023}, \cc{ZIF-71 \cite{ortiz2014b,sun2021}}, Cu$_2$(tebpz) \cite{grosu2016,anagnostopoulos2020,chorazewski2021}, and, although less stable \cite{grosu2015a}, ZIF-67 and other ZIF-8 derivatives \cite{grosu2015a,khay2016,mortada2018,sun2021} have been tested.  Overall, MOFs have opened a new era in HLS applications \cite{tortora2021,sun2021} and have stimulated interesting fundamental questions \cite{johnson2023,amayuelas2023}, but their impact could be much broader if more stable hydrophobic reticular materials were synthesised. 

Typical energy applications of HLS rely on compression/decompression cycles over a sufficiently broad pressure range to trigger intrusion and (in most cases) boiling \cc{(Fig.~\ref{fig:int-ext}a)}; depending on the hysteresis, these cycles dissipate or store and release energy. For instance, these cycles could be used to dissipate mechanical vibrations \cite{eroshenko2007,iwatsubo2007,suciu2008} and shocks \cite{sun2021} or as ``molecular springs'' \cite{eroshenko2002,trzpit2007,grosu2016}. However, mechanical energy is not the only possible form in which energy is exchanged in HLS \cite{grosu2017}, see Fig.~\ref{fig:int-ext}a. Energy was harnessed by isobaric low-temperature cycles in Cu$_2$(tebpz), realising a compact and efficient thermal actuator \cite{chorazewski2021}. Triboelectric effects could also be a future direction to obtain directly electrical energy from intrusion/boiling cycles \cite{grosu2017,lowe2019}.

\subsection{High Performance Liquid Chromatography}
\label{sec:HPLC}

High-Performance Liquid Chromatography (HPLC) is a popular analytical technique used to separate components based on the different affinity of the analytes dissolved in the mobile phase for the stationary phase which fills the HPLC column. In reversed-phase liquid chromatography (RPLC)\cite{colin1977}  a polar mobile phase, typically mixtures of water and organic solvents, in which analytes with different degrees of hydrophobicity are dissolved, is flowed at high pressure through 
a non-polar stationary phase made of a hydrophobic material, typically mesoporous silica gels with different functionalisations \cite{cruz1997}, analogous to the materials discussed in Sec.~\ref{sec:energy} for HLS. The requirement to decrease the environmental impact of solvents, together with specific applications to separate very polar compounds, pushes towards the adoption of increasing fractions of water in the mobile phase, which is the greenest possible solvent \cite{dembek2020}. 
The framework of Sec.~\ref{sec:phase} and the examples of Sec.~\ref{sec:energy} clearly suggest that highly aqueous mobile phases may boil at such conditions, making the nanopores unavailable to the analytes (``retention loss''). This phenomenon has indeed been known in the chromatographic community as ``phase collapse'' and correctly ascribed to boiling (``dewetting'') only recently \cite{walter1997,walter2005}.
In accord with the physical insights presented in Sec.~\ref{sec:phase}, recent systematic investigations \cite{gritti2019,gritti2020} showed that 1) low salt concentration do not significantly influence dewetting; 2) the presence of dissolved nitrogen in the aqueous eluent can have some effect on boiling; 3) the dewetting process is strongly dependent on the column temperature; 4) the pore characteristics (porosity, connectivity, and pore size distribution) affect dewetting, in line with what  reported \cc{elsewhere} \cite{amabili2019}.

Protocols to measure retention losses by dewetting were recently published \cite{gritti2019}, which allow to draw a parallel with the boiling phenomena described in Sec.~\ref{sec:phase}. Indeed the experience accumulated with HLS and the related theoretical advancements could be beneficial for the HPLC community to better understand and control dewetting, while the HLS community could benefit from the advanced control over the surface functionalisations which has been developed along the years for RPLC.
For example, it has been recently suggested that fine details of the functionalisation can lead to large differences in both intrusion and boiling \cite{rasera2023}: chain length and grafting densities, within the range used in applications, can change the boiling pressure by more than $40$~MPa. Controlling such phenomena is of interest both for the HLS community and for the HPLC one. 
Several advancements are underway, including the formulation of best practices \cite{gritti2020} to avoid dewetting in RPLC, but a quantitative understanding is still elusive and the related multiscale computational tools are to be fully developed; several interesting phenomena are to be explored, including how the presence of analytes at the surface affects boiling. 

\subsection{Hydrophobic Gating in Ion Channels and Nanopores}
\label{sec:hygate}
\begin{figure}[h!t]
    \centering
    \includegraphics[height=0.3\textwidth]{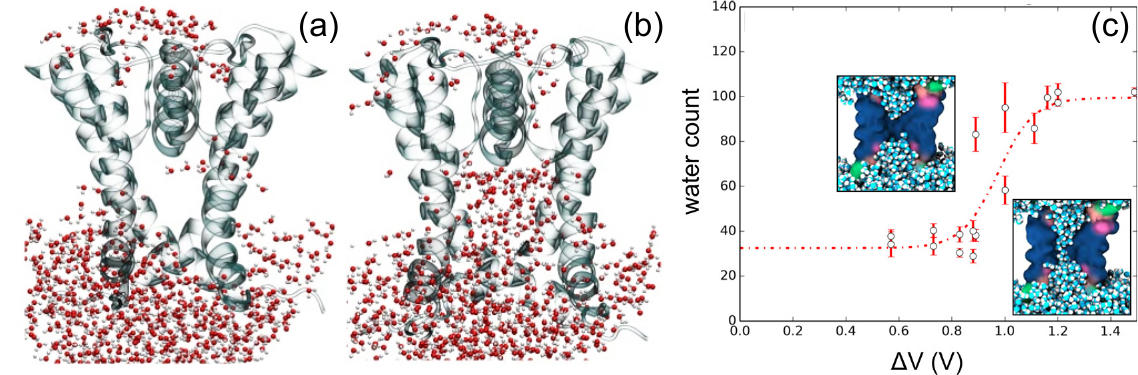}
\caption{Hydrophobic gating in the BK channel: closed (a) and open (b) states show different levels of water occupation. Reproduced from Jia et al., Nat. Commun. 9, 3408 (2018). Copyright 2018 Authors. c) Electrowetting in a biomimetic channel allows control of boiling in nanopores by acting on the voltage difference $\Delta V$ across the membrane. Reproduced from Trick et al., ACS Nano 11, 1840 (2017). Copyright 2017 American Chemical Society.}
    \label{fig:ionchannel}
\end{figure}
Ion channels are transmembrane proteins that enable the transport of ions with their hydration shell across the hydrophobic cellular membrane \cite{hille2001}. The typical structure of ion channels involves the presence of a selectivity filter, which allows the conduction of  some types of ions only, a pore, which can host the ion together with several water molecules, and the gate, which is generally in charge of switching on or off the ionic currents \cite{doyle1998}. Ion channels open and close in response to different stimuli, including voltage, concentration, mechanical stress, and temperature \cite{hille2001} -- a tempting but still unexplored analogy to meta-MOFs \cite{coudert2019}. Several gating mechanisms \cc{are known}, which typically involve the steric occlusion of the gate on the intracellular side \cite{black2021}. However, some channel structures, in which the gate is sufficiently open to allow for ion conduction, display instead the extremely low conductivities characteristic of closed channels \cite{rao2019}; this is one of the signatures of hydrophobic gating, in which the flow of water and ions is blocked by the formation of a nanoscale bubble \cite{aryal2015,rao2018} rather than by a steric constriction. 
\cc{Figure~\ref{fig:ionchannel} shows the closed (a) and open (b) states of the big potassium channel BK, from which it is clear that boiling phenomena can switch the ion flux}. 
Indeed, the theoretical considerations reported in Sec.~\ref{sec:phase}, show that the presence of hydrophobic aminoacids, together with (sub)nanoscale confinement, could give rise to boiling at physiological conditions \cite{roth2008,giacomello2020}. 

Molecular dynamics simulations of ion channels currently serve as an invaluable bridge between protein structure, dynamics, and function \cite{dror2012,guardiani2022}, especially in view of the lack of simple direct measurements of bubble formation. Molecular dynamics showed that the then available structure of the bacterial mechanosensitive channel of small conductance MscS exhibited hydrophobic gating, thus being non conductive and functionally closed \cite{anishkin2004}. Other notable examples of channels in which hydrophobic gating occurs are  MscL 
\cite{anishkin2010}, the nicotinic acetylcholine receptor \cite{beckstein2006,chiodo2017},  GLIC  \cite{zhu2012a},  BK \cite{jia2018} (Fig.~\ref{fig:ionchannel}a-b), and  CRAC \cite{guardiani2021}. Also classical density functional theory calculations have shown the formation of a low  density region in the gate of a model KcsA channel \cite{gussmann2017}. Euristic approaches have been developed to identify hydrophobic gates in the expanding database of channel structures \cite{rao2019}.

The concept of hydrophobic gating was first formulated starting from simple model nanopores \cite{beckstein2003,roth2008}, similar to those used in other contexts, e.g., HLS \cite{tinti2017}.
This suggests again that other biological nanopores, simpler than ion channels, could display boiling within their lumen. Indeed, it has been recently reported that the toxin FraC could be engineered by mutating two aminoacids in the pore constriction with two hydrophobic ones to produce hydrophobic gating: free-energy molecular dynamics showed that, at low pH, a wet, conductive state and an empty, non-conductive one are present  which was confirmed by electrophysiology measurements at different pH \cite{paulo2023c}. While the formation of bubble is generally considered detrimental in nanopore sensing applications \cite{smeets2006b,smeets2008}, it could prove useful to embed some functionalities of voltage-gated ion channels into artificial or hybrid nanopores \cite{powell2011,smirnov2011,paulo2023c}.
\cc{Figure~\ref{fig:ionchannel}c shows an engineered $\beta$-barrel nanopore in which hydrophobic gating could be controlled by applying a suitable voltage across the membrane.}

\section{Future perspectives}
\label{sec:perspectives}

In the following, selected topics of emerging interest for nanopore boiling are briefly touched upon, to show the ebullience of the topic and to \cc{hopefully} inspire new experimental, theoretical, and computational endeavours.

\subsection{Angstroscale effects}
Boiling at the nanoscale demonstrates the complementarity of experiments, theory, and computations. On the one hand, experiments are the privileged means of discovering new phenomenology and the final benchmark of quantitative science. On the other hand, experiments at the (sub)nanoscale often lack the resolution or the control over parameters to be self-standing: interpretation is required by theory or by  simulation, which provide a microscopic connection to macroscopic observables. Theory is the crucial reduction tool that allows to test hypotheses, scalings, and to exclude some of the many intervening factors from the intricate nanoscale jungle. At the same time, classical theories are challenged by boiling in nanopores, calling from microscopic investigations of non-continuum, angstroscale effects, in addition to structural and chemical heterogeneities. Atomistic simulations represent the tool of choice to investigate these aspects and to bridge experiments at different scales: for instance, experiments on HLS typically quantify macroscopic volumes and pressures (or heat fluxes), but require a microscopic interpretation to disclose intrusion or boiling mechanisms \cite{amayuelas2023} or the contribution of mesoscale quantities, as line tension \cite{guillemot2012a,tinti2017}. Even more strikingly, ion channels research can typically rely on structures  obtained by electron microscopy and functional information from electrophysiology experiments, two pieces of information at very different scales in time and space; the bridge is often provided by molecular dynamics which has the suitable resolution 
\cite{dror2012,guardiani2022}.

Transport properties in subnanoconfinement have been the object of increasing attention owing to  the unprecedented control over nanotubes and ``angstroslits''
\cite{kavokine2021,you2022}; reported results challenge the current understanding of angstrom scale flows, e.g., wall slip which is dependent on electronic properties of the nanotube rather than on structural ones \cite{secchi2016}. 
In the context of boiling in nanopores, microporous materials afford control of molecular details and pose similar angstroscale questions challenging continuum but also mesoscale concepts. 
Importantly, the foundational concepts of contact angle and radius entering  Kelvin-Laplace eq.~\eqref{eq:lcap} break down at these scale, leaving room to locally varying chemical and geometrical heterogeneities; this is probably at the origin of the typically low intrusion pressures of MOFs \cite{ortiz2013,grosu2015a,grosu2015b,grosu2016} as compared to the macroscopic expectations based on their radii.
Mesoscale concepts as bending rigidities and line tension maybe also fail at scales in which individual hydrogen bonds matter, which are directed by the structure of the reticular materials \cite{coudert2021,johnson2023,paulo2023b}.
Indeed, the structure of microporous materials  could be leveraged to tune the macroscopic behaviour of HLS, e.g., subnanometric connections between micropores could impact the intrusion  \cite{bushuev2022,bushuev2022b} or boiling \cite{paulo2023b} pressures, due to their capability to control hydrogen bonds across the main pores \cite{johnson2023,paulo2023b} \cc{see, e.g., Fig.~\ref{fig:nanoscale}}.
Similarly, biological nanopores exhibiting hydrophobic gating lend themselves as a flexible platform  to explore the fundamentals of boiling at the angstroscale, by exploiting the panoply of biophysical techniques, e.g., performing systematic and targeted point mutations within the pore \cite{lucas2021,paulo2023c}; this introduces the next future perspective.

\subsection{Quantitative biology and bioinspiration}
Biological nanopores pose several challenges to the current understanding of boiling, which have been touched upon in the previous sections, including chemical and structural complexity, flexibility, presence of ions and electrical fields. The development of theoretical and simulation tools is helping to promote a quantitative understanding of boiling in biological nanopores \cite{roth2008,giacomello2020} and, at the same time, to provide the computer aided  tools to design nanopore applications \cite{paulo2023c}.
This is a small part of the grand endeavour of reaching a quantitative description of biological phenomena \cite{noble2002}, which involves the concurrence of multiple disciplines and multiscale models. 

If the quantitative understanding of biological phenomena has a clear and yet challenging program, even more promising and unexplored in the field of boiling in nanopores is the opposite process of bioinspiration, i.e., learning from biological phenomena, using biological concepts, and realising hybrids. 
In ion channels hydrophobic gating is just a part of a more complex mechanism of signal transduction and actuation, which in turn enables more complex functions, e.g., the action potential in neurons \cite{hodgkin1952}. One extraordinary property of ion channels is coupling ion selectivity, which is so sophisticated to filter out smaller ions (e.g., sodium) while being conductive for larger potassium \cite{doyle1998}, with high conductivity \cite{horn2014}. Imitating such capability would be crucial to develop advanced nanoporous membranes for reverse osmosis or other applications \cite{werber2016,zhang2022,you2022}.
Another property of ion channels that is currently under the spotlight in HLS research is flexibility, which is emerging as a means to tailor the boiling conditions \cite{lowe2019,tortora2021,caprini2023}; in proteins this idea is brought to the extreme, with substantial conformational changes occurring during gating, which enable to switch  between different boiling conditions \cite{guardiani2021}, which in turn correspond to different conduction properties.
Even more profoundly, such conformational changes happen in response to a signal which is sensed elsewhere in the channel, e.g., at the voltage sensing domain \cite{hille2001}. Such allosteric response mechanisms could inspire new active routes to switch at a distance the conductive properties  of a microporous material \cite{coudert2019} when a signal is sensed. 
Finally, through the environment-dependent tuning of their conductive properties, ion channels orchestrate more complex functions, as the action potential \cite{hodgkin1952}, which is a basic route to transmit information across the human body. In this biomimetic direction, hydrophobic gating has been proposed as an elementary mechanism to mimic the capabilities of ion channels at neuromorphic computing; the underlying physics is voltage-regulated boiling which tunes the conductivity of the nanopore, allowing for the emergence of memory \cite{paulo2023c}.

\subsection{Emerging applications}

\begin{figure}[h!t]
    \centering
    \includegraphics[height=0.5\textwidth]{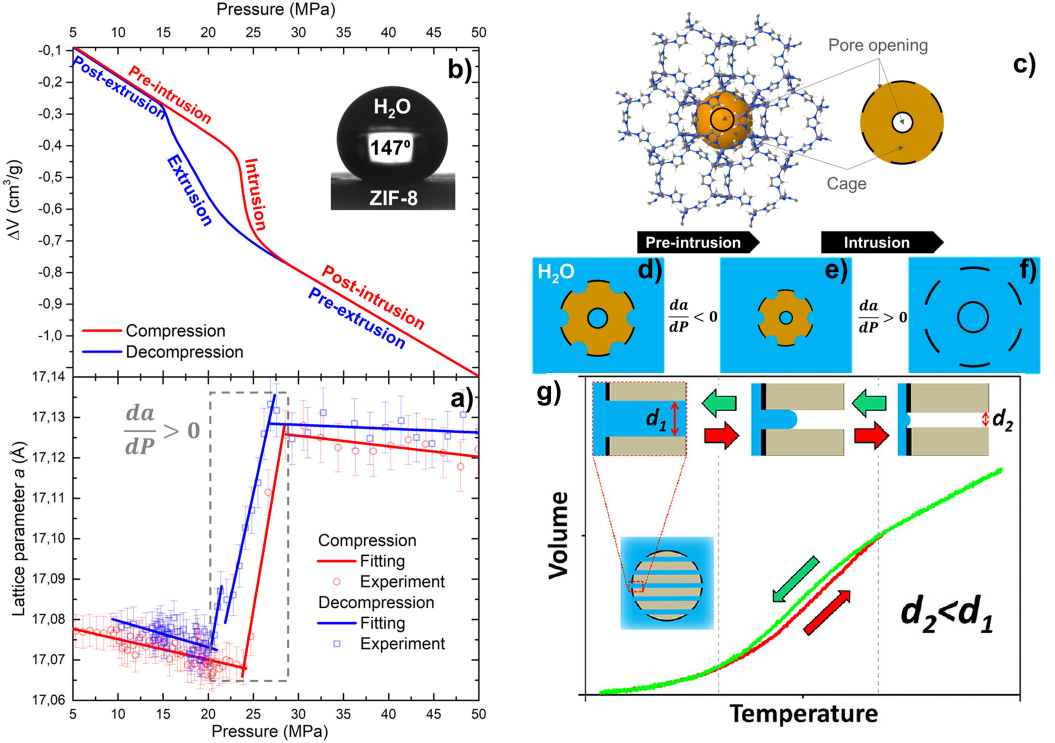}
\caption{Negative compressibility of ZIF-8 during compression (red) and decompression (blue) is related to intrusion and boiling, respectively. a) Changes in the lattice parameter $a$ as a function of pressure measured by in situ neutron scattering experiments for a system comprising ZIF-8 and D$_2$O. b) Pressure-volume isotherm  and contact angle of the ZIF-8 + D$_2$O system. c) ZIF-8 structure highlighting the internal cavity and the pore openings connecting the cages; d-f) illustration of the  response of ZIF-8 to compression before (d-e) and after  water intrusion (f). Here,  blue  represents water, while  black indicates the dimensions of the cavity.
Panels a-f are reprinted with permission from Tortora et al., Nanoscale 12, 22698 (2020). Copyright 2021 American Chemical Society.
g) Isobaric thermal cycle for the Cu$_2$(tebpz) MOF immersed in water displays effective thermal-to-mechanical energy conversion: starting from wet pores and increasing temperature (red line), boiling occurs where the slope increases. Cooling the system (green line) allows it to return to the initial state with minimal hysteresis. Reprinted with permission from Chorazewski et al., ACS Nano 15, 9048 (2021). Copyright 2021 American Chemical Society.
\label{fig:NC}}
\end{figure}
In addition to the more consolidated uses exposed in Sec.~\ref{sec:impact}, the exquisite control of phase transitions, in particular boiling, achieved by microporous materials has the potential to enable a number of new applications. Energy absorption is one of the earliest proposed utilization of HLS \cite{suciu2003,eroshenko2007}; MOFs, however, allowed to push the capabilities of HLS to realise shock-absorbers that are reusable and capable of absorbing more energy the \cc{faster} the impact \cite{lowe2019,sun2021}.
Unconventional negative compressibility \cite{nicolaou2012} properties of MOFs ensuing from the coupling of boiling in nanopores and elastic properties (Fig.~\ref{fig:NC}a-f) has been exploited in CO$_2$ sensing applications \cite{anagnostopoulos2020} and for proposing pressure-sensitive valves for nanofluidic applications \cite{tortora2021}. At the origin of negative compressibility are elastocapillary phenomena, which allow the material to shrink during hydrostatic decompression, because of the formation of menisci within the pores, or to expand during hydrostatic compression because of their suppression \cite{caprini2023}, \cc{see Fig.~\ref{fig:NC}d-f}. 
Negative thermal expansion was shown to give rise to thermal-to-mechanical energy conversion with a remarkable efficiency of ca. $30\%$ over a rather limited range of temperatures ($30\,^\circ$C to $90\,^\circ$C) \cite{chorazewski2021}. \cc{Figure~\ref{fig:NC}g shows the operating principle of such a device based on the Cu$_2$(tebpz) MOF: a thermal cycle is performed in which the temperature-induced reduction in the pore diameter triggers boiling when the temperature is raised, leading to a monotonic increase in the system volume, which could be used for thermal actuation}. 
Finally, triboelectrification during intrusion/boiling cycles \cite{grosu2017,lowe2019} shows promise to convert directly mechanical vibrations, e.g., coming from car suspensions, to electrical energy.

\section{Conclusions}
\label{sec:conclusions}
The physical perspective adopted in Sec.~\ref{sec:phase} allowed the dissection of nanopore boiling in different contributions: starting from the macroscopic ones due to confinement size and hydrophobicity expressed by Kelvin-Laplace eq.~\eqref{eq:lcap} in Sec.~\ref{sec:fundamentals}, passing through the contributions characteristic of the nanoscale (e.g., curvature and line effects, Sec.~\ref{sec:nanoscale}), and arriving to the effects intrinsic to the pore structure (e.g., pore connectivity, flexibility, heterogeneties, Sec.~\ref{sec:intrinsic}) and of the fluid and extrinsic characteristics (temperature, cycle time, electric field, etc., Sec.~\ref{sec:extrinsic}). These physical insights may be useful to devise new solutions and quantitative tools (Sec.~\ref{sec:perspectives}) to meet the technological requirements of energy materials (Sec.~\ref{sec:energy}) and chromatographic columns (Sec.~\ref{sec:HPLC}), e.g., tuning the intrusion/boiling hysteresis in HLS or designing HPLC stationary phases or protocols which allow to work with green solvents avoiding dewetting issues. 
Finally, a strong physical basis is key to understand the biological phenomenon of hydrophobic gating (Sec.~\ref{sec:hygate}), which is fundamentally the same as boiling in artificial nanopores, but with all the complexity which was progressively unrolled in Sec.~\ref{sec:phase} occurring at once and in a mutually dependent way.

\begin{acknowledgments}
The author thanks Y. Grosu and S. Meloni for thoughtful discussions \cc{and G. Paulo for critically reading the manuscript}.
This research is part of a project that has received funding from the European Research Council (ERC) under the European Union’s Horizon 2020 research and innovation programme (grant agreement No. 803213).
\end{acknowledgments}

\bibliography{bibliography}

\end{document}